\documentclass[apj, iop]{emulateapj}
\usepackage{url}
\usepackage{amsmath}

\newcommand{\tna}{\tablenotemark{a}}
\newcommand{\tnb}{\tablenotemark{b}}
\newcommand{\tnc}{\tablenotemark{c}}
\newcommand{\NH}{\ensuremath{N_\mathrm{H}}}

\newcommand{\NHo}{\ensuremath{N_{\mathrm{H}^0}}}
\newcommand{\NDo}{\ensuremath{N_{\mathrm{D}^0}}}
\newcommand{\NOo}{\ensuremath{N_{\mathrm{O}^0}}}
\newcommand{\Ho}{\ensuremath{\mathrm{H}^0}}
\newcommand{\Do}{\ensuremath{\mathrm{D}^0}}
\newcommand{\lya}{\ensuremath{\rm Ly\alpha}}

\newcommand{\kms}{\rm km~s\ensuremath{^{-1}\,}}
\newcommand{\cmm}{\rm cm\ensuremath{^{-2}}}
\newcommand{\cmmm}{\rm cm\ensuremath{^{-3}}}
\newcommand{\msun}{\ensuremath{\rm M_\odot}}
\newcommand{\msunyr}{\ensuremath{\rm M_{\odot}\;{\rm yr}^{-1}}}
\newcommand{\ergscmmHz}{\ensuremath{{\rm erg~s}^{-1}~{\rm cm}^{-2}~{\rm Hz}^{-1}}}

\newcommand{\Alp}{\ensuremath{\rm Al^+}}
\newcommand{\AlII}{\hbox{{\rm Al}{\sc \,ii}}}

\newcommand{\CII}{\hbox{{\rm C}{\sc \,ii}}}
\newcommand{\CIIs}{\hbox{{\rm C}{\sc \,ii}$^*$}}

\newcommand{\CIV}{\hbox{{\rm C}{\sc \,iv}}}
\newcommand{\DI}{\hbox{{\rm D}{\sc \,i}}}
\newcommand{\FeII}{\hbox{{\rm Fe}{\sc \,ii}}}
\newcommand{\HI}{\hbox{{\rm H}{\sc \,i}}}
\newcommand{\MgI}{\hbox{{\rm Mg}{\sc \,i}}}
\newcommand{\MgII}{\hbox{{\rm Mg}{\sc \,ii}}}

\newcommand{\NII}{\hbox{{\rm N}{\sc \,ii}}}
\newcommand{\NV}{\hbox{{\rm N}{\sc \,v}}}
\newcommand{\OI}{O{\sc\,i}}
\newcommand{\Oo}{\ensuremath{\rm O^0}}
\newcommand{\OVI}{\hbox{{\rm O}{\sc \,vi}}}

\newcommand{\SiII}{\hbox{{\rm Si}{\sc \,ii}}}

\newcommand{\SiIV}{\hbox{{\rm Si}{\sc \,iv}}}

\begin{document}

\lefthead{Metal-Poor, Cool Gas in the Circumgalactic Medium of a $z=2.4$ Star-Forming Galaxy}\righthead{N.~H.~M.~Crighton et al.}


\title{Metal-Poor, Cool Gas in the Circumgalactic Medium of a $z=2.4$ Star-Forming Galaxy: Direct Evidence for Cold Accretion?\altaffilmark{*}}


\author{Neil H. M. Crighton\altaffilmark{1},  Joseph F. Hennawi\altaffilmark{1} \& J. Xavier Prochaska\altaffilmark{1,2}}

\altaffiltext{*}{Data and code used for this paper are available at \protect\url{http://github.com/nhmc/z2.4CGM}}
\altaffiltext{1}{Max-Planck-Institut f\"ur Astronomie, K\"onigstuhl}
\altaffiltext{2}{Department of Astronomy and Astrophysics, 
  UCO/Lick Observatory; University of California, 1156 High Street, Santa Cruz, CA 95064; xavier@ucolick.org}


\begin{abstract}

In our current galaxy formation paradigm, high-redshift galaxies are
predominantly fuelled by accretion of cool, metal-poor gas from the
intergalactic medium.  Hydrodynamical simulations predict that this
material should be observable in absorption against background
sightlines within a galaxy's virial radius, as optically thick
Lyman-limit systems (LLSs) with low metallicities.  Here we report the
discovery of exactly such a strong metal-poor absorber at an impact
parameter $R_\perp=58$~kpc from a star-forming galaxy at $z=2.44$.
Besides strong neutral hydrogen ($\NHo=10^{19.50\pm0.16}~\cmm$) we
detect neutral deuterium and oxygen, allowing a precise measurement of
the metallicity: $\log_{10}(Z/Z_\odot)=-2.0\pm0.17$, or
$(7-15)\times10^{-3}$ solar. Furthermore, the narrow deuterium
linewidth requires a cool temperature $<20,000$~K. Given the striking
similarities between this system and the predictions of simulations,
we argue that it represents the direct detection of a high redshift
cold-accretion stream. The low-metallicity gas cloud is a single
component of an absorption system exhibiting a complex velocity,
ionization, and enrichment structure. Two other components have
metallicities $>0.1$ solar, ten times larger than the metal-poor
component. We conclude that the photoionized circumgalactic medium
(CGM) of this galaxy is highly inhomogeneous: the majority of the gas
is in a cool, metal-poor and predominantly neutral phase, but the
majority of the metals are in a highly-ionized phase exhibiting weak
neutral hydrogen absorption but strong metal absorption. If such
inhomogeneity is common, then high-resolution spectra and detailed
ionization modelling are critical to accurately appraise the
distribution of metals in the high-redshift CGM.

\end{abstract}


\keywords{galaxies: halos --- galaxies: evolution --- galaxies: high-redshift --- quasars: absorption lines --- intergalactic medium}



\section{Introduction}

It is now recognized that the physical processes by which galaxies
accrete, expel and recycle baryons are of central importance to galaxy
evolution \citep[e.g.][]{Dave12}. The current theoretical paradigm
emphasizes a `cold-mode' of accretion \citep{Birnboim03, Keres05} that
fuels the high star formation rates (SFRs) observed in the early
universe. This cool ($T\sim 10^4$~K), accreting gas is predicted to
produce observable signatures as high column density, low metallicity
absorption systems \citep{Fumagalli11, FaucherGiguere11, Stewart11,
  vandeVoort12_dla, Kimm11, Goerdt12, Nelson13} in the spectra of
background sightlines which pass through the circumgalactic medium
(CGM) of high redshift galaxies. However, there is little direct
observational evidence for the cold-accretion phenomenon. Instead,
metal-enriched outflows are commonly observed from $z\sim2-3$
Lyman-break galaxies (LBGs) as blue-shifted absorption features in the
galaxies' spectra \citep{Pettini01_LBG,Steidel10}. Energetic
`feedback' in some form is required to regulate star formation in
numerical simulations, which is necessary to match observables such as
the stellar luminosity function \citep{Oppenheimer10} and galaxy
rotation curves \citep{Stinson13_EarlyFeedback}, but the distances to
which these outflows travel and their energetics are poorly
constrained. Precise measurements of the enrichment and physical
conditions of the CGM are thus crucial both to test the cold accretion
picture and to characterize how far outflowing material travels.

The best technique currently available to measure metallicities is by
analyzing individual absorption line profiles from gas in the
CGM. This technique has been used extensively to measure metallicities
of high column density systems over a broad redshift range
\citep[e.g.][]{Steidel90_LLS,Lehner13}. However, the relationship
between the absorbing gas and galaxies in these studies is often
unclear, making them difficult to interpret in the context of galaxy
formation models. The cold accretion phenomenon may persist to lower
redshifts in lower mass haloes, and several observational studies have
discovered low metallicity ($Z/Z_\odot=0.01-0.1$) absorbers in the
neighbourhood of $z<1$ galaxies
\citep[e.g.][]{Chen05_dla,Tripp05,Ribaudo11,Kacprzak12,Churchill12}. But
measurements of CGM enrichment near vigorously star-forming galaxies
at $z\sim2.5$, during the peak of the cosmic SFR and where cold
accretion is predicted to occur over a wide range of halo masses, have
proven to be much more challenging.

In this letter we analyze an echelle spectrum of a background QSO at
an impact parameter of 58~kpc from a foreground star-forming galaxy
discovered by \citet{Rudie12}\footnote{All distances given are proper,
  assuming $H_0=70\,\kms\,\mathrm{Mpc}^{-1}$, $\Omega_M=0.3$ and
  $\Omega_\Lambda=0.7$.}. The high signal-to-noise ratio (${\rm
  S/N\sim 90}$), high-resolution (${\rm FWHM}\sim8.8~\kms$) spectrum
combined with the rare detection of the deuterium isotope of hydrogen,
enable us to effectively disentangle absorption produced by neutral,
metal poor gas from that of more highly ionized, metal enriched
gas. The deuterium detection allows us to precisely measure the
metallicity of a metal-poor CGM absorber, and we show that it exhibits
many of the characteristics expected for cold-mode accretion.

\section{Observations and Analysis}

\subsection{The Background QSO and Foreground Galaxy}

The bright background QSO~J1444535$+$291905 ($g=16.4$) at redshift
$z=2.660$ has been observed using the HIRES echelle spectrograph on
Keck by several groups from 1995 to 2009. We obtained the spectra from
the Keck archive and reduced them with
\textsc{makee}\footnote{\url{http://www.astro.caltech.edu/~tb/makee/}}. The
final combined spectrum covers wavelength ranges $3100$ to
$5950$~\AA\ (resolution $8.8\,\kms$) and $7350$ to
$9760$~\AA\ (resolution $6.7\,\kms$).

The foreground galaxy's redshift ($z_{gal}=2.4391$) and impact
parameter from the background QSO ($R_{\perp}=58$~kpc) are given by
\citet[][their figure 12]{Rudie12}. The redshift uncertainty is
$160-180$~\kms, which arises when rest-frame UV transitions such as
\lya\ emission and metal absorption are used to determine the
intrinsic redshift \citep{Rakic11}. Rudie et al. do not publish
further details on the galaxy, but such galaxies typically have a
dust-extinction corrected SFR of 30~\msunyr\ \citep{Erb06_sfr}, a
gas-phase ISM metallicity of $Z/Z_\odot\sim0.8$ \citep{Erb06_Z} and a
dark matter halo mass $5\times10^{11}~\msun$ \citep{Bielby13}. This
halo mass implies a virial radius of 75~kpc \citep{Bryan98}, and thus
the projected separation from the background QSO sightline is
comparable to the virial radius.

We note that \citet{Rudie13} indicates a second star-forming galaxy is
present at $R_\perp=50$~kpc within $250~\kms$ of the first galaxy, but
does not publish the spectroscopic redshift. We adopt $R_\perp=58$kpc
for our analysis, but our conclusions remain unchanged if a second
galaxy exists at an even closer impact parameter.

\subsection{The Absorption System at $z=2.44$}

\begin{figure*}
\begin{minipage}[p]{\linewidth}
\centering \includegraphics[width=0.6\columnwidth]{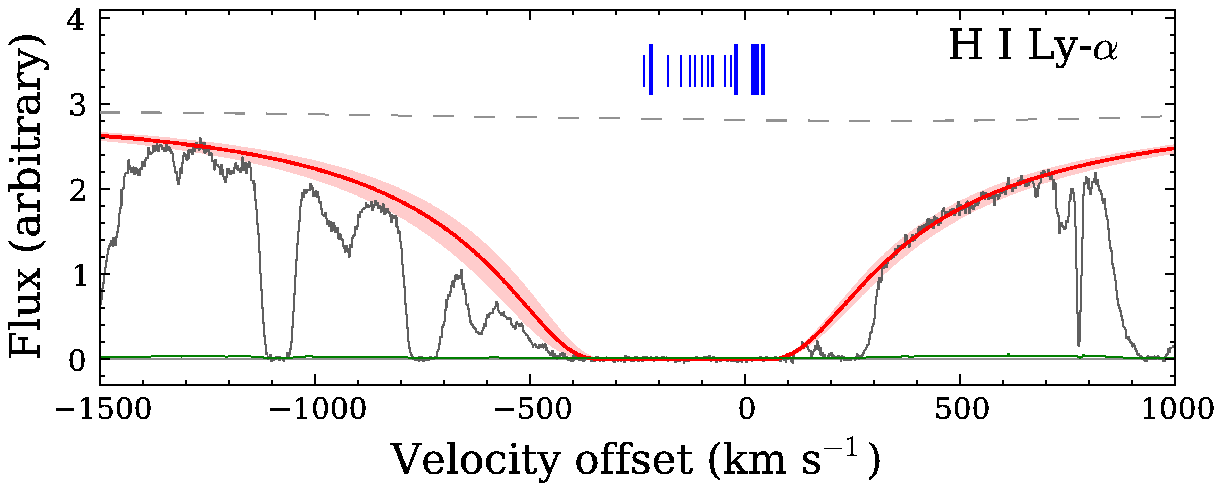}
\end{minipage}\\[5mm]
\begin{tabular}{c c}
\includegraphics[width=0.95\columnwidth, scale=1]{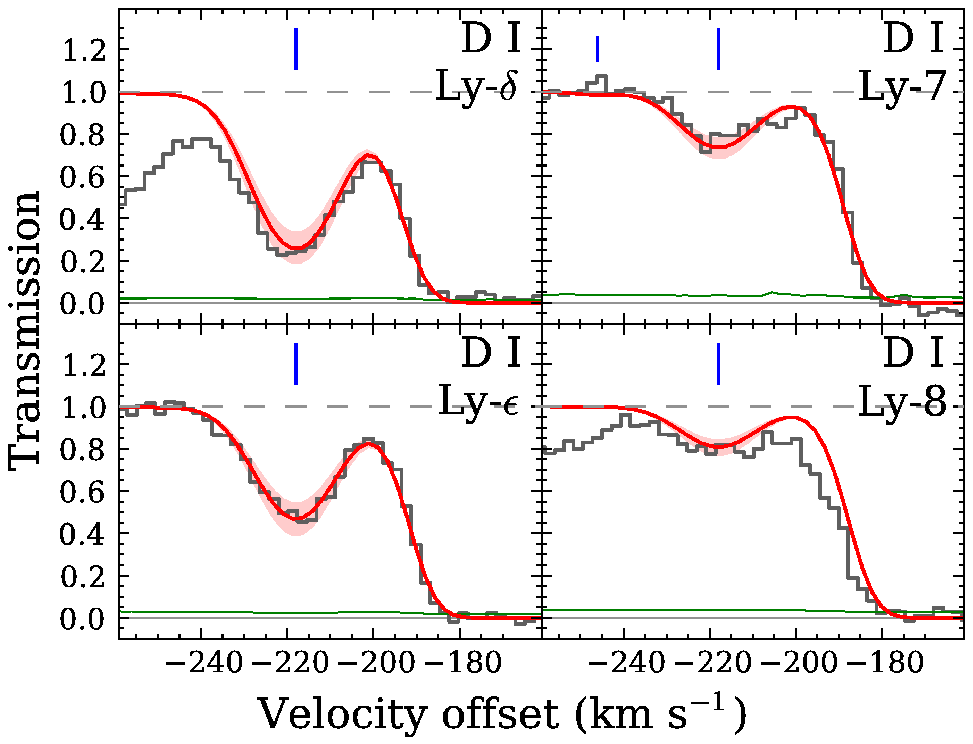}&
\includegraphics[width=0.95\columnwidth,scale=1]{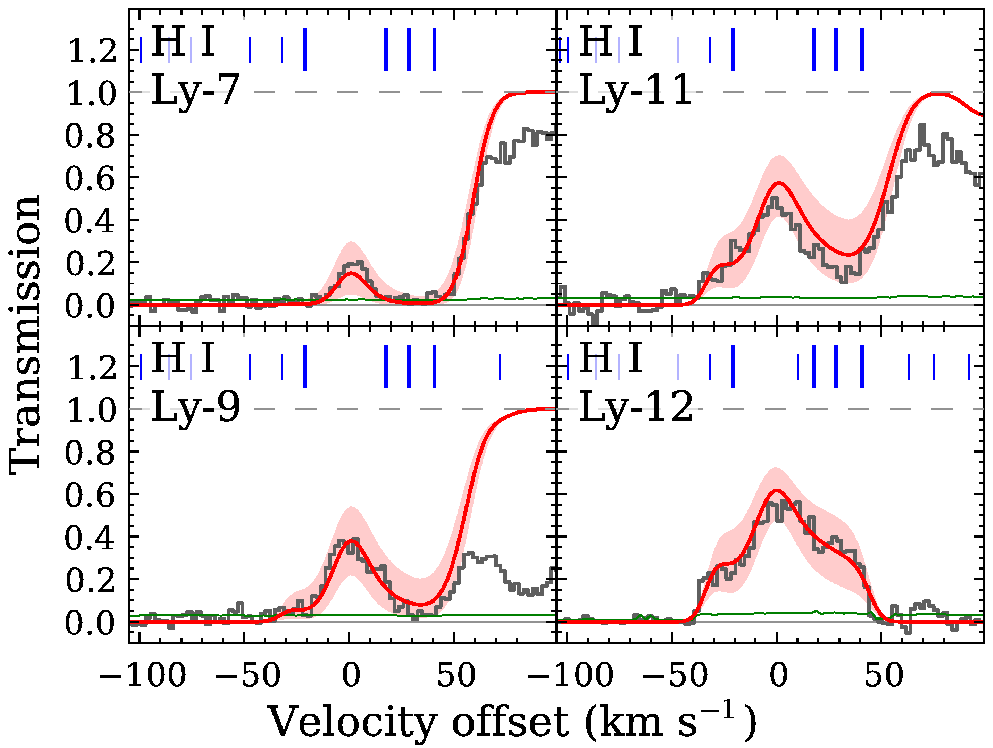}
\end{tabular}
\caption{\label{f_HI} HIRES spectra showing absorption from the D and
  partial LLS components, marked with long ticks. Shorter ticks show
  other low-ionization metal components without tight constraints on
  \NHo. Velocities are relative to $z=2.4391$, and our
  best-fitting model is plotted over the data. Panels show
  \HI\ \lya\ (top), and deuterium absorption from the D component (left) and
  \HI\ for the partial LLS components (right). Shaded regions around
  the model show the effect of our adopted column density
  uncertainties for \Ho\ and \Do.}
\end{figure*}

\begin{figure*}
\centering \includegraphics[width=1.99\columnwidth]{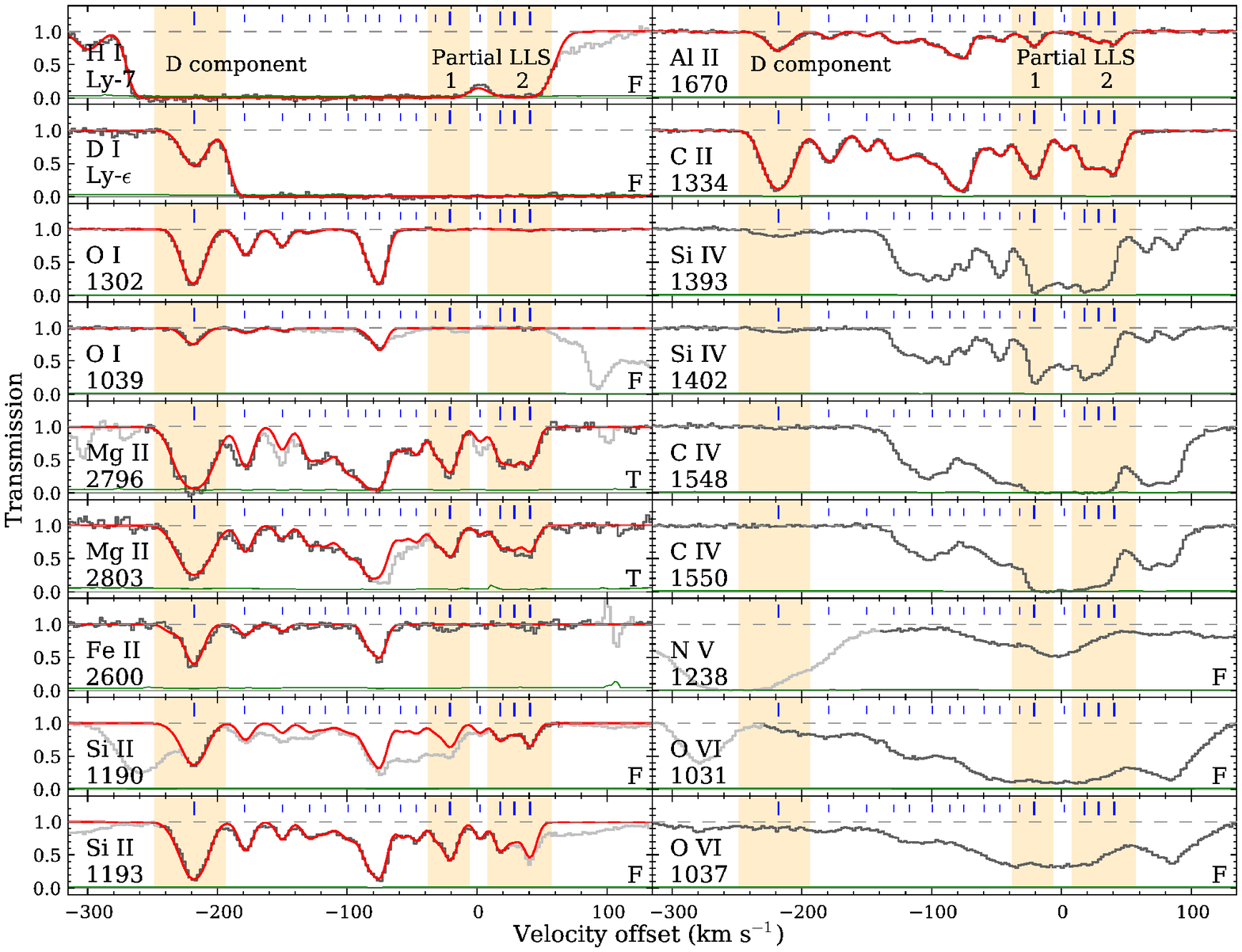}
\caption{\label{f_metals}  HIRES spectra showing metal transitions for
  the $z=2.44$ absorption system. Velocities are relative to
  $z=2.4391$, and our best-fitting model is plotted over the data. `F'
  indicates there is blending with \lya\ forest absorption; `T'
  blending with telluric lines. Shading marks regions used to measure
  AOD column densities, and greyed out lines show regions heavily
  blended with unrelated forest or telluric absorption.}
\end{figure*}

The HIRES spectrum reveals a sub-damped absorption system, originally
discovered by \citep{Simcoe02}, that is coincident with the redshift
of the foreground galaxy. Our Voigt profile fit to the damping wings
implies a total neutral hydrogen column density
$\NHo=10^{19.85\pm0.1}$~\cmm\ (Figure~\ref{f_HI}, top). Absorption
from low and high-ionization ionic metal line transitions is present
in 16 distinct components spread across
$320\,$\kms\ (Figure~\ref{f_metals}). The total $\NHo$ is
well-constrained, but as we will demonstrate, the metallicity varies
by an order of magnitude among the components. Fortuitously, we can
also constrain \NHo\ for several \emph{individual} components.

We measured column densities in the system by fitting Voigt profiles
with
\textsc{vpfit}\footnote{\url{http://www.ast.cam.ac.uk/~rfc/vpfit.html}},
and by using the apparent optical depth \citep[AOD,][]{Savage91}
method. Voigt profile fits assess the degree to which a single
kinematic model can explain all the observed absorption:
low-ionization transitions were used to define the main velocity
components, and we found that a single velocity structure matches all
the low-ionization species well. The resulting models are shown in
Figures~\ref{f_HI} \& \ref{f_metals}. The higher ionization
transitions \SiIV, \CIV\ and \OVI\ do not follow the same velocity
structure as the low-ions, indicating there is likely a contribution
from a hotter or more highly-ionized phase.  Similar offsets between
low and high-ions have been observed in other high-$z$ absorbers
\citep[e.g.][]{Fox07}. Table~\ref{t_one} gives Voigt profile
parameters and our adopted column densities for each component. Our
Voigt profile fits to both the high-ions and low-ions show no evidence
for partial covering or unresolved saturated components.

\subsection{Components for which Abundances Can be Precisely Measured}

We identify three kinematic regions for which the neutral hydrogen
column \NHo, and thus the metallicity, can be precisely measured. The
first is at $-218$~\kms\ relative to the galaxy redshift, where
absorption from neutral deuterium is detected at Ly-$\gamma$,
Ly-$\delta$, Ly-$\epsilon$ and Ly-8. Deuterium detections in QSO
absorbers are rare, as a measurement of the weak D absorption requires
a narrow linewidth, high \NHo, and the absence of blending with weaker
neighbouring \lya\ forest lines \citep{Webb91}.  Such detections are
valuable however, because a precise D/H measurement constrains the
cosmic baryon density according to big bang nucleosynthesis
theory. For the D component we detect, $\NDo=10^{14.93\pm0.06}\cmm$
and $b(\Do)=10.60\pm0.53~\kms$, which are consistent with the expected
column density and width of the associated \Ho, estimated from the
damping wings and the Lyman series, assuming thermal broadening and
the cosmic D/H ratio inferred from the WMAP cosmic baryon density
\citep{Keisler11}. This width is much narrower than is typical for a
\lya\ forest line with the $\NHo\approx10^{15}\cmm$ necessary to
imitate \DI, which demonstrates that this is not an interloping
\lya\ forest line.

We can use the D/H ratio implied by the cosmic baryon density to make
a precise measurement of \NHo\ in this component. \citet{Pettini12}
find $\log_{10}({\rm D/H})=-4.585\pm0.02$ in QSO absorbers. The
standard deviation in these QSO values (0.15) is larger than that
expected from the measurement errors, and represents real scatter due
to an unidentified astrophysical effect and/or underestimated
systematic errors in the D/H analyses. We conservatively assume the
scatter represents a physical variation and take $\log_{10}({\rm
  D/H})~=-4.585\pm0.15$.  \Do/\Ho\ is the same as D/H, and therefore
$\log_{10}\NHo=19.50\pm0.16$ for this component.

Crucially, we also detect \OI\ in this component, which allows a
direct measurement of the metallicity from O$^0/$H$^0$. For $\NHo
\gtrsim 10^{19}~\cmm$, the ion ratio O$^0/$H$^0$ is a robust
metallicity indicator, because the ionization potentials of \Oo\ and
\Ho\ are very similar and the two species exchange electrons
\citep{Field71}, and O does not deplete strongly onto dust grains
\citep{Jenkins09}.  We find [O/H]~$=-2.00\pm0.17$, which is comparable
to the background metallicity of the IGM at this redshift
\citep{Schaye03, Simcoe04}.

The remaining components for which we can accurately measure \NHo\ are
partial Lyman limit systems at $\sim-20$ and $\sim+30$~\kms (Partial
LLS 1 \& 2, see Figure~\ref{f_metals}). These show unsaturated
\HI\ absorption at the Ly-10, Ly-11 and Ly-12 transitions, which imply
$\NHo=10^{16.30\pm0.20}$ (Partial LLS 1) and $10^{16.43\pm0.30}~\cmm$
(Partial LLS 2). Since ionization corrections are significant at these
\NHo, we estimate the metallicity of these components using the
photoionization models described in the following section.

\subsection{Photoionization Modelling}
\label{s_cloudy}

\begin{figure*}
\centering
\begin{tabular}{c}
\includegraphics[scale=0.7]{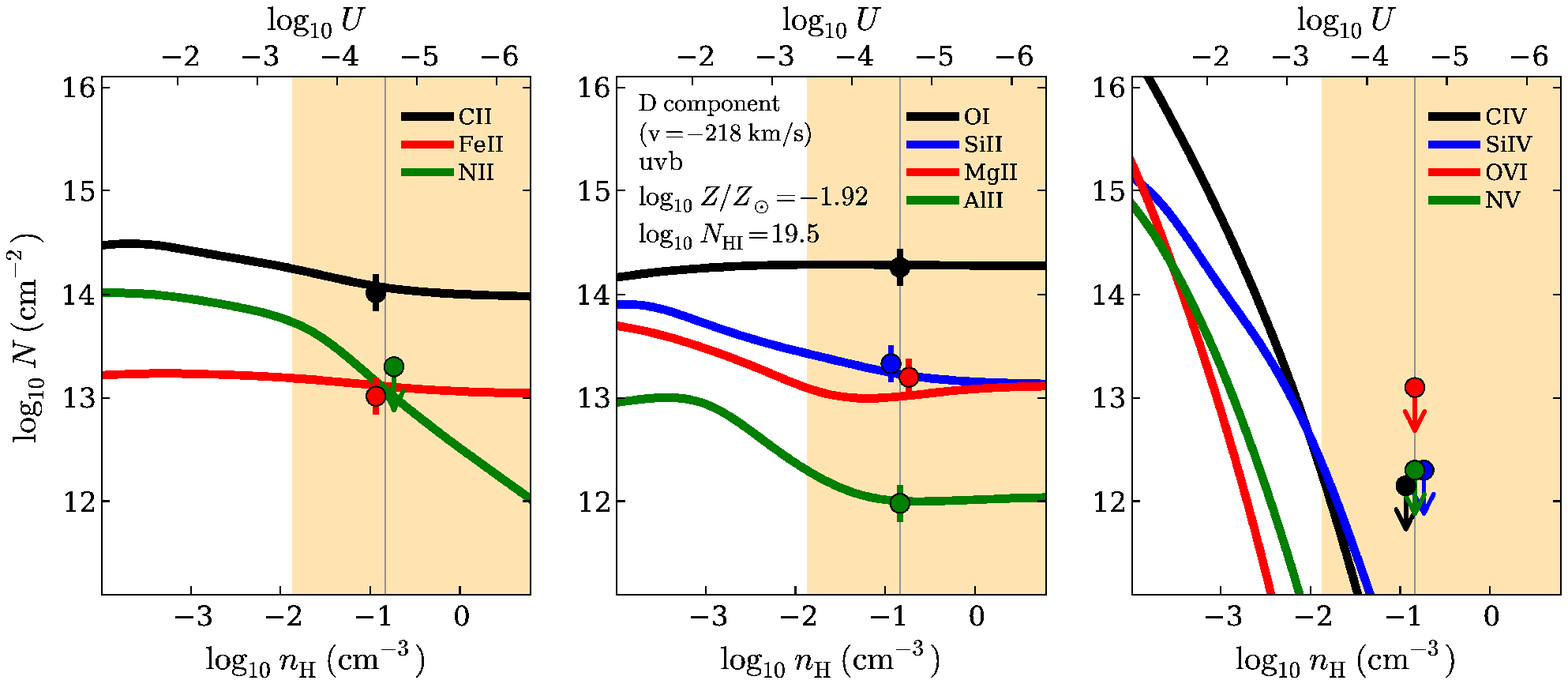}\\[2mm]
\includegraphics[scale=0.7]{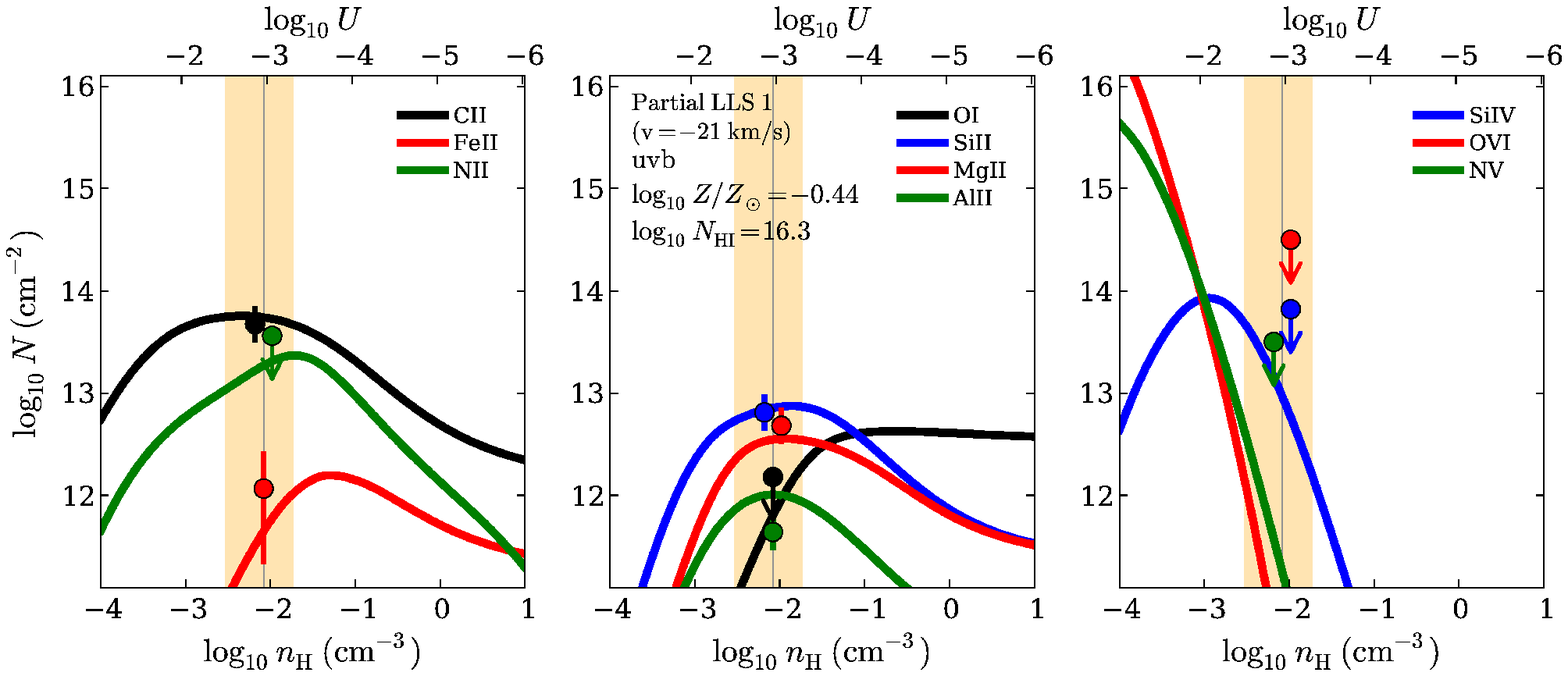}\\[2mm]
\includegraphics[scale=0.7]{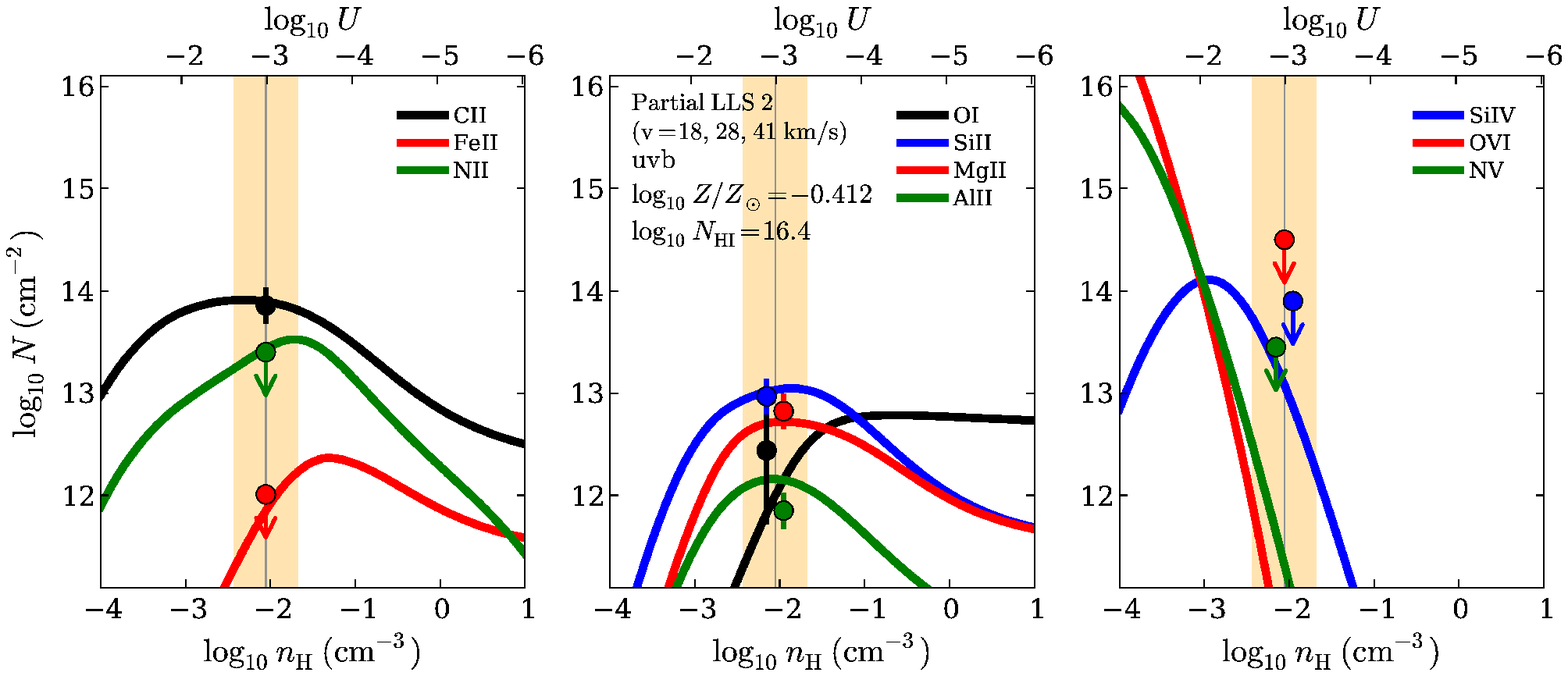}
\end{tabular}
\caption{\label{f_cloudy2} \textsc{cloudy} models for the D
  component (top), Partial LLS 1 (middle) and 2 (bottom). Curves show
  the model predictions, and points show the observed values and
  uncertainties at the best-fitting $U$ values, sometimes offset for
  clarity. The shaded region shows the $U$ range consistent with the
  data. Error bars include the observed column density uncertainty and
  a 0.15 dex uncertainty to account for relative abundance
  variations. }
\end{figure*}

We use \textsc{cloudy} photoionization models \citep[version
  8.01][]{Ferland98} to predict the metal ion column densities for the
D and partial LLS components given a metallicity and ionization
parameter $U$ (defined as the ratio of the densities of ionizing
photons to hydrogen atoms). All models assume solar abundance ratios,
no dust, and an ionizing spectrum due to the ambient UV background
\citep{Haardt12}\footnote{we also consider a spectrum dominated by a
  starburst galaxy, but it does not significantly affect our
  results.}. We use a maximum likelihood technique to find the
best-fitting metallicity and ionization parameter for each component
(Crighton et al., in preparation) and their uncertainties. Column
densities for low-ions together with upper limits on high-ions are
used to constrain the models. Figure~\ref{f_cloudy2} shows the models
for the D component and partial LLS components, each at the
best-fitting metallicity for that component.

For all three components we find that the UV background models provide
a good match to the data. For the partial LLS components there is a
minor discrepancy for $N_{\rm \Alp}$, which is $0.3-0.4$ dex lower
than predicted by the models. This could be due to uncertainties in
the shape of the ionizing spectrum, although using a
starburst-dominated ionizing spectrum makes the disagreement more
severe. We believe this discrepancy is likely due to an Al
underabundance relative to a solar abundance pattern
\citep[e.g.][]{Rolleston03}. The predicted N$^+$ column is close to
the observed upper limit for Partial LLS 2, which may indicate that
nitrogen is underabundant relative to solar. A nitrogen underabundance
is often observed in damped systems at a similar redshift
\citep[e.g.][]{Prochaska02a}. For all other ionic species there is
excellent agreement between the observed column densities and model
predictions.

\subsection{Physical Conditions in the Absorption System}
\label{s_phys}

We have used two independent methods to show that the metallicity of
the D component is $\sim1/100$ solar: $Z/Z_\odot=10^{-2.00\pm0.17}$
using O$^0/$H$^0$, and $Z/Z_\odot=10^{-1.92\pm0.18}$ from the
\textsc{cloudy} models using all the metal transitions except
\OI. The width of the \DI\ line requires a gas temperature
$<20,000$~K, conservatively assuming thermal broadening such that
$b=\sqrt{2kT/m_{\rm D}}$. From the models we find $\log_{10}U=-4.2\pm
0.95$, which implies that H is $\sim75\%$ neutral
($\NH\sim10^{19.6}~\cmm$) and has a temperature of $\sim10,000$~K, in
agreement with $b$(\DI). [Fe/O]~$=-0.10\pm0.09$, therefore there is no
evidence for strong dust depletion, as is expected from the low
metallicity. The maximum $U$ permitted ($10^{-3.2}$) corresponds to a
neutral fraction of $25\%$ and $n_\mathrm{H}=10^{-1.8}~\cmmm\times
F_\nu^{\,912}/(3\times10^{-21}~\ergscmmHz)$, where $F_\nu^{\,912}$ is
the magnitude of the ionizing spectrum at 912~\AA. This implies a
cloud thickness $\lesssim3$~kpc.

In contrast to the D component, Partial LLS 1 and 2 are metal enriched
and highly ionized. Both LLS have similar conditions: the models yield
$Z/Z_\odot=10^{-0.44\pm0.31}$ ($n_\mathrm{H}\approx10^{-2}~\cmmm\times
F_\nu^{\,912}/(3\times10^{-21}~\ergscmmHz)$) and $\log_{10}U=-3.10\pm
0.15$, implying that less than $1\%$ of H is neutral
($\NH\sim10^{19}~\cmm$). Using a starburst-dominated ionizing spectrum
allows a lower $Z/Z_\odot=10^{-0.8\pm0.32}$, but results in a 0.6-0.7
dex disagreement between the observed and predicted
$N_{\rm\Alp}$. Therefore we prefer the UVB-only models, but our main
conclusions are unchanged if we were to instead use a
starburst-dominated spectrum. A further difference between the D and
partial LLS components is the presence of absorption from higher
ionization gas (\SiIV, \CIV, \OVI) over the same velocity range as the
low-ions. While some \SiIV\ components align with the low-ions,
overall the high-ions have a different velocity structure, and are
much stronger than expected from our photoionization modelling of the
low-ions. Very broad components are also present in \OVI\ (see
Figure~\ref{f_metals}). Thus these high-ions may trace another gas
phase that is shock-heated and collisionally ionized \citep[see
  e.g.][]{Simcoe02}.

Finally, we calculate [O/H] for the entire system, including all
low-ion components, assuming \Oo\ is associated with the bulk of
the \Ho\ and that ionization corrections are negligible. This gives
[O/H]$_{\rm total}=-1.89\pm0.11$. Therefore all the \OI\ components,
spanning $\sim140~\kms$, must have a metallicity similar to the D
component.

\section{Discussion}

\begin{figure}
\centering
\includegraphics[width=0.9\columnwidth]{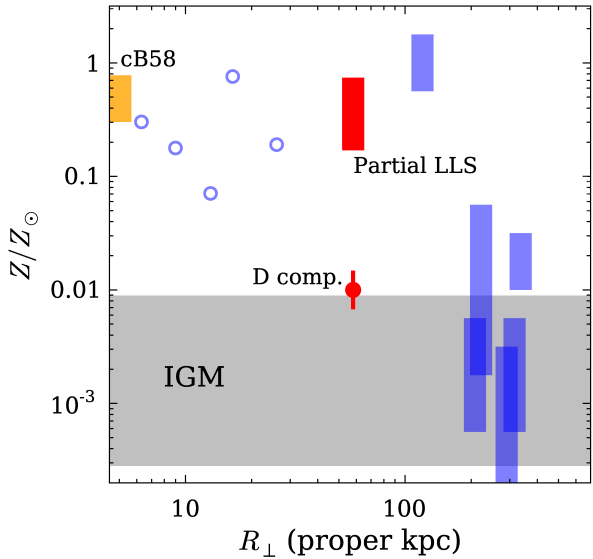}
\caption{\label{f_met} Metallicity of gas associated with $z\sim2.5$
  LBGs as a function of impact parameter. The ISM metallicity measured
  in the lensed LBG cB58 is shown at $R_{\perp}=4$~kpc. Other
  $R_\perp<30$~kpc measurements are for galaxies found by targeted
  searches around known neutral, damped systems. The D component and
  partial LLSs are shown by a red circle and red
  rectangle. Measurements at larger separations are from
  \citet{Simcoe06}.}
\end{figure}

Figure~\ref{f_met} shows the current state of gas-absorption
metallicities measured around $z\sim2.5$ LBGs. Enriched gas with
$Z/Z_{\odot}\gtrsim0.3$ is found in the partial LLS component in this
paper at $R_\perp=58$~kpc, for the galaxy-absorber pair at
$R_\perp=110$~kpc from \citet{Simcoe06}, and in the ISM of the
gravitationally-lensed LBG cB58 \citep{Pettini02_cB58}, shown at
$R_{\perp}\sim4$~kpc. Systems at larger separations, also from Simcoe
et al., approach the metallicity of the surrounding IGM, suggesting
that enriched gas is common in the CGM up to
$R_\perp\sim100$~kpc. Also shown are measurements from highly-neutral
damped systems with nearby galaxies \citep{Krogager12,
  Bouche13}. However, we caution that these systems were pre-selected
based on their absorption properties, and so they may not be
representative of the typical circumgalactic medium.
The metallicity of the D component is more than an order of magnitude
lower than the typical ISM metallicity of LBGs \citep{Pettini02_cB58},
indicating that this gas has not been recently stripped or ejected in
a wind from the galaxy. Could it be produced by a cold-mode accretion
flow? It is close to the virial radius\footnote{We estimate the chance
  of a random sub-DLA falling within 250~\kms\ of the galaxy redshift
  at $<2.2\%$ based on the 12 QSO-galaxy pairs searched to find this
  system.}, has a high column density ($10^{19.50\pm0.16}~\cmm$), low
metallicity ($10^{-2.00\pm0.17}$), cool temperature
($20,000\mathrm{K}\ll T_{virial}\approx 10^6\,$K), and characteristic
size $\lesssim 3$ kpc, which are all properties of inflowing streams
expected from simulations
\citep[e.g.][]{Fumagalli11}. \citet{Stewart11} predict that accreting
streams have high angular momentum and often co-rotate with the
galaxy's disk, producing offsets of up to $100$~\kms\ from the
systemic galaxy redshift. The D component is at
$-218\pm180$~\kms\ from the nearby galaxy, which is compatible with
such an offset. All of the \OI\ components in the absorber, covering a
velocity range of $\sim140$~\kms, also have low metallicities. Such
velocity separations are seen by Stewart et al. when a sightline
intersects multiple accreting streams. Therefore, in light of all the
evidence above, we believe that this absorber represents the best
existing candidate for a cold-mode accretion flow at high redshift.

The CGM around this galaxy is highly inhomogeneous: the D and partial
LLS components have metallicities that differ by at least a factor of
ten. These different metallicities are reminiscent of the bimodal
metal distribution seen in $z<1$ Lyman limit systems \citep{Lehner13},
which Lehner et al. argue results from physically distinct processes:
metal-poor accreting gas, and metal-enriched gas stripped from
galaxies or ejected in large-scale winds. Indeed, the two components
in this system show further differences besides the metallicity: the D
component is neutral and has very weak high-ion absorption, whereas
the partial LLSs are highly ionized and have strong associated
high-ions. These high-ions are not produced by the same ionizing
spectrum that can reproduce the low-ions, and they may be caused by
shock-heated, collisionally-ionized gas that is expected in a
supernovae-driven outflowing wind \citep[e.g.][]{Simcoe06}. This is
also consistent with the hydrodynamical simulations of \citet{Shen13},
which indicate that enriched outflowing and metal-poor inflowing
phases of the CGM are mixed kinematically.

If such a bimodality is common, as is implied by analyses of other
$z\sim3$ sub-DLAs \citep[e.g.][]{Prochter10}, it has fundamental
implications for interpreting observations of the CGM. Using the
\NH\ values inferred from photoionization modelling and the component
metallicities, we conclude that the highly ionized partial LLSs
contain {\it most} of the total metal mass in this system,
significantly more than the low-metallicity component which dominates
the total \HI\ absorption. Consider the case of this same system
observed at a lower resolution, or of a composite spectrum which
averages the absorption of many such systems
\citep[e.g.][]{Steidel10}. In this case, the metal absorption for
individual components can no longer be matched to their respective
\HI, and we must measure the metallicity for the system as a whole. We
showed in \S\ref{s_phys} that doing this results in a low metallicity,
[O/H]$\approx-1.9$. Thus the high metallicity partial LLSs would
escape detection. An even more interesting case to consider is if the
low-metallicity component had been more highly ionized, which would
occur if it was illuminated by a stronger radiation field, or if it
had a lower total \NH, reducing its ability to self-shield. To explore
this case we ran photoionization models for a range of $U$ values,
using $\NHo=10^{17}\cmm$, and with other properties identical to the D
component. These show that for a total $\NH\le10^{20}\cmm$,
characteristic of cold streams \citep{Fumagalli11}, the strongest
low-ion metal transition lying outside the
\lya\ forest\footnote{Higher ionization states \SiIV\ and \CIV\ cannot
  be used to detect this gas as they also arise in another co-spatial
  gas phase.}, \CII\ $\lambda1334$, always has a column density
$<10^{13.1}\cmm$. Such a weak line approaches the detection limit even
for echelle spectra. In low-resolution or composite spectra it would
be easily swamped by stronger absorption from any kinematically
nearby, metal-enriched gas, and the total metallicity inferred for the
system would be dominated by metal-rich components. Therefore a large
reservoir of metal-poor gas may be missed by existing studies of the
$z\sim2.5$ CGM that use only low-resolution or composite
spectra. Further studies employing high-resolution spectra of
background QSOs near foreground galaxies are essential to accurately
assess the metal content of the CGM.

By creating a large sample of similar absorption systems where
metallicity measurements can be made at small impact parameters, we
can realise the exciting possibility of mapping the enrichment
distribution of the high-redshift CGM. Both our group and others
\citep[e.g.][]{Rudie12} are pursuing programs to assemble such a
sample.

\acknowledgments

We thank J.O'Meara for a preliminary reduction of J1444535$+$291905,
and the referee, whose comments improved the paper.

\begin{table}
\renewcommand{\arraystretch}{1.2}
%\addtolength{\tabcolsep}{-0.5pt}
\footnotesize
%\scriptsize
\begin{center}
\caption{\label{t_one} Absorption parameters}
\begin{tabular}{rcccccc}
\tableline
 Ion & $z$ &  $b$ & $\log_{10} N$ (VPFIT)  & $\log_{10} N$ (AOD) & Adopted $\log_{10} N$ & AOD Transition      \\
     &     & (km~s$^{-1}$)& (cm$^{-2}$)    & (cm$^{-2}$) & (cm$^{-2}$) & ($\lambda_{\rm rest}$, \AA)   \\
\tableline
%\vspace{0.2cm}
\multicolumn{7}{l}{{\bf D component} (one component at $-218\,\kms$, AOD range $-248.5$ to $-193.5\,\kms$)} \\[\smallskipamount]
\HI      & 2.436599(01)& $<18$            & $19.50 \pm 0.16$          &        --                     & $19.50 \pm 0.16$             &        \\
\DI      &             & $10.60\pm 0.53$  & $14.93 \pm 0.02$          & $14.976^{+0.061}_{-0.064}$    & $14.93 \pm 0.06$             & 937    \\
\AlII    &             & 10.42\tna        & $11.96 \pm 0.02$          & $11.980^{+0.071}_{-0.075}$    & $11.980^{+0.071}_{-0.075}$   & 1670   \\
\CII     &             & $11.4 \pm 0.2$   & $14.04 \pm 0.01$          & $14.014^{+0.018}_{-0.018}$    & $14.014^{+0.018}_{-0.018}$   & 1334   \\
\CIIs    &             &                  & $<13.1$                   & $13.863^{+0.018}_{-0.018}$    & $<13.1$                      & 1335.7 \\
\CIV     &             &                  & $<12.15$                  & $12.115^{+0.377}_{-0.862}$    & $<12.15$                     & 1548   \\
\FeII    &             & 7.40\tna         & $12.98 \pm 0.04$          & $13.015^{+0.088}_{-0.082}$    & $13.015^{+0.088}_{-0.082}$   & 2600   \\
\MgI     &             &                  & --                        & $<13.9$                       & $< 13.9$                     & 1683   \\
\MgII    &             & $13.7 \pm 0.8$   & $13.20 \pm 0.02$          & $13.198^{+0.102}_{-0.094}$    & $13.198^{+0.102}_{-0.094}$   & 2803   \\
\NII     &             &                  & $<13.3$                   & $< 13.901$                    & $<13.3$                      & 1084   \\
\NV      &             &                  & $<12.3$                   & $< 12.558$                    & $<12.3$                      & 1242   \\
\OI      & 2.436590(01)\tnb&$8.6 \pm 0.13$&$14.303\pm0.005$           & $14.261^{+0.014}_{-0.013}$    & $14.261^{+0.047}_{-0.013}$   & 1302   \\
\OVI     &             &                  & $<13.1$                  & $< 13.323$                     & $<13.1$                      & 1031   \\
\SiII    &             &$10.43\pm0.17$\tna& $13.36 \pm 0.006$         & $13.329^{+0.049}_{-0.051}$    & $13.329^{+0.049}_{-0.051}$   & 1526   \\
\SiIV    &             &                  & $<12.30$                  & $12.323^{+0.118}_{-0.146}$    & $<12.30$                     & 1393   \\[\smallskipamount]
\multicolumn{7}{l}{{\bf Partial LLS 1}  (one component at $-21\,\kms$, AOD range $-38$ to $-6\,\kms$)} \\[\smallskipamount]
\HI      & 2.438862(04)& $14.4\pm1.7$     &$16.30 \pm 0.20 $          &   --                          &  $16.30 \pm 0.20$            &        \\
\AlII    &             & 4.91\tna         &$11.63 \pm 0.03$           &  $11.642^{+0.087}_{-0.107}$   &  $11.642^{+0.087}_{-0.107}$  & 1670   \\
\CII     &             & 7.35\tna         &$13.66 \pm 0.02$           &  $13.674^{+0.015}_{-0.015}$   &  $13.674^{+0.015}_{-0.015}$  & 1334   \\
\CIIs    &             &                  &$<11.9$                    &  $< 12.021$                   &  $< 11.9$                    & 1335.7 \\
\CIV     &             &                  &$>14.3$\tnc                &  $>14.29$\tnc                     &  $>14.3$                     & 1550   \\
\FeII    &             &                  &$<12.08$                   &  $12.067^{+0.335}_{-0.712}$   &  $12.067^{+0.335}_{-0.712}$  & 2600   \\
\MgI     &             &                  & --                        &  $<13.8$                      &  $< 13.8$                    & 1683   \\
\MgII    &             & 5.17\tna         &$12.61 \pm 0.04$           &  $12.681^{+0.084}_{-0.087}$   &  $12.681^{+0.084}_{-0.087}$  & 2796   \\
\NII     &             &                  &$<13.56$                   &  $<13.787$                    &  $<13.56$                    & 1084   \\
\NV      &             &                  &$<13.5$                    &  $<13.541$                    &  $<13.5$                     & 1238   \\
\OI      &             &                  &$<12.18$                   &  $<12.349$                    &  $<12.18$                    & 1302   \\
\OVI     &             &                  &$<14.5$                    &  $14.290^{+0.019}_{-0.018}$   &  $<14.5$                     & 1037   \\
\SiII    &             & $4.81\pm0.27$\tna&$12.83\pm0.01$             &  $12.811^{+0.096}_{-0.101}$   &  $12.811^{+0.096}_{-0.101}$  & 1526   \\
\SiIV    &             &                  &$<13.82$                   &  $13.480^{+0.019}_{-0.019}$   &  $<13.82$                    & 1402   \\[\smallskipamount]
\multicolumn{7}{l}{{\bf Partial LLS 2}  (three components at $18, 28, 41\,\kms$, AOD range $8$ to $57.5\,\kms$)}    \\[\smallskipamount]
\HI      &             &                  & $16.43 \pm 0.30$          &  --                           &  $16.43 \pm 0.30$          &       \\
\AlII    &             &                  & $11.859^{+0.032}_{-0.034}$&  $11.851^{+0.087}_{-0.104}$   &  $11.851^{+0.087}_{-0.104}$& 1670  \\
\CII     &             &                  & $13.880^{+0.025}_{-0.026}$&  $13.856^{+0.015}_{-0.015}$   &  $13.856^{+0.015}_{-0.015}$& 1334  \\
\CIIs    &             &                  & $<12.1$                   &  $< 12.114$                   &  $< 12.1$                  & 1335.7\\
\CIV     &             &                  & $>14.4$\tnc               &  $>14.4$\tnc                  &  $>14.4$                   & 1550  \\
\FeII    &             &                  & $<12.01$                  &  $< 12.307$                   &  $< 12.01$                 & 2600  \\
\MgI     &             &                  & --                        &  $<13.9$                      &  $< 13.9$                  & 1683  \\
\MgII    &             &                  & $12.865^{+0.041}_{-0.046}$&  $12.823^{+0.076}_{-0.076}$   &  $12.823^{+0.076}_{-0.076}$& 2796  \\
\NII     &             &                  & $<13.4$                   &  $13.316^{+0.072}_{-0.089}$   &  $<13.4$                   & 1084  \\
\NV      &             &                  & $<13.45$                  &  $13.358^{+0.037}_{-0.041}$   &  $<13.45$                  & 1238  \\
\OI      &             &                  & $12.395^{+0.184}_{-0.327}$&  $12.439^{+0.383}_{-0.698}$   &  $12.439^{+0.383}_{-0.698}$& 1302  \\
\OVI     &             &                  & $<14.5$                   &  $14.311^{+0.020}_{-0.020}$   &  $<14.5$                   & 1037  \\
\SiII    &             &                  & $13.080^{+0.016}_{-0.017}$&  $12.968^{+0.098}_{-0.114}$   &  $12.968^{+0.098}_{-0.114}$& 1526  \\
\SiIV    &             &                  & $<13.9$                   &  $13.609^{+0.019}_{-0.019}$   &  $<13.9$                   & 1402
\end{tabular}
\tablenotetext{1}{\scriptsize $b$ value is tied to other species assuming purely thermal broadening.}
\tablenotetext{2}{\scriptsize A single-component Voigt profile fit to the \OI\ line is offset $\sim0.8\,\kms$ from the \CII\ line. This may indicate unresolved component structure, but we do not expect it to significantly affect \NOo.}
\tablenotetext{3}{\scriptsize Saturated components.}
%\vspace{-0.8cm}

\tablecomments{
  Errors on the AOD column densities include uncertainties
  in the continuum and zero levels. High-ion upper limits are
  calculated by measuring the highest column density Voigt profile
  consistent with the data, assuming thermal broadening.}

\end{center}
\end{table}

%% %% The following command ends your manuscript. LaTeX will ignore any text
%% %% that appears after it.


\begin{thebibliography}{48}
\expandafter\ifx\csname natexlab\endcsname\relax\def\natexlab#1{#1}\fi

\bibitem[{{Bielby} {et~al.}(2013){Bielby}, {Hill}, {Shanks}, {Crighton},
  {Infante}, {Bornancini}, {Francke}, {H{\'e}raudeau}, {Lambas}, {Metcalfe},
  {Minniti}, {Padilla}, {Theuns}, {Tummuangpak}, \& {Weilbacher}}]{Bielby13}
{Bielby}, R., {Hill}, M.~D., {Shanks}, T., {Crighton}, N.~H.~M., {Infante}, L.,
  {Bornancini}, C.~G., {Francke}, H., {H{\'e}raudeau}, P., {Lambas}, D.~G.,
  {Metcalfe}, N., {Minniti}, D., {Padilla}, N., {Theuns}, T., {Tummuangpak},
  P., \& {Weilbacher}, P. 2013, \mnras, 430, 425

\bibitem[{{Birnboim} \& {Dekel}(2003)}]{Birnboim03}
{Birnboim}, Y., \& {Dekel}, A. 2003, \mnras, 345, 349

\bibitem[{{Bouch{\'e}} {et~al.}(2013){Bouch{\'e}}, {Murphy}, {Kacprzak},
  {P{\'e}roux}, {Contini}, {Martin}, \& {Dessauges-Zavadsky}}]{Bouche13}
{Bouch{\'e}}, N., {Murphy}, M.~T., {Kacprzak}, G.~G., {P{\'e}roux}, C.,
  {Contini}, T., {Martin}, C., \& {Dessauges-Zavadsky}, M. 2013, ArXiv e-prints

\bibitem[{{Bryan} \& {Norman}(1998)}]{Bryan98}
{Bryan}, G.~L., \& {Norman}, M.~L. 1998, \apj, 495, 80

\bibitem[{{Chen} {et~al.}(2005){Chen}, {Kennicutt}, \& {Rauch}}]{Chen05_dla}
{Chen}, H.-W., {Kennicutt}, Jr., R.~C., \& {Rauch}, M. 2005, \apj, 620, 703

\bibitem[{{Churchill} {et~al.}(2012){Churchill}, {Kacprzak}, {Steidel},
  {Spitler}, {Holtzman}, {Nielsen}, \& {Trujillo-Gomez}}]{Churchill12}
{Churchill}, C.~W., {Kacprzak}, G.~G., {Steidel}, C.~C., {Spitler}, L.~R.,
  {Holtzman}, J., {Nielsen}, N.~M., \& {Trujillo-Gomez}, S. 2012, \apj, 760, 68

\bibitem[{{Dav{\'e}} {et~al.}(2012){Dav{\'e}}, {Finlator}, \&
  {Oppenheimer}}]{Dave12}
{Dav{\'e}}, R., {Finlator}, K., \& {Oppenheimer}, B.~D. 2012, \mnras, 421, 98

\bibitem[{{Erb} {et~al.}(2006{\natexlab{a}}){Erb}, {Shapley}, {Pettini},
  {Steidel}, {Reddy}, \& {Adelberger}}]{Erb06_Z}
{Erb}, D.~K., {Shapley}, A.~E., {Pettini}, M., {Steidel}, C.~C., {Reddy},
  N.~A., \& {Adelberger}, K.~L. 2006{\natexlab{a}}, \apj, 644, 813

\bibitem[{{Erb} {et~al.}(2006{\natexlab{b}}){Erb}, {Steidel}, {Shapley},
  {Pettini}, {Reddy}, \& {Adelberger}}]{Erb06_sfr}
{Erb}, D.~K., {Steidel}, C.~C., {Shapley}, A.~E., {Pettini}, M., {Reddy},
  N.~A., \& {Adelberger}, K.~L. 2006{\natexlab{b}}, \apj, 647, 128

\bibitem[{{Faucher-Gigu{\`e}re} \& {Kere{\v s}}(2011)}]{FaucherGiguere11}
{Faucher-Gigu{\`e}re}, C.-A., \& {Kere{\v s}}, D. 2011, \mnras, 412, L118

\bibitem[{{Ferland} {et~al.}(1998){Ferland}, {Korista}, {Verner}, {Ferguson},
  {Kingdon}, \& {Verner}}]{Ferland98}
{Ferland}, G.~J., {Korista}, K.~T., {Verner}, D.~A., {Ferguson}, J.~W.,
  {Kingdon}, J.~B., \& {Verner}, E.~M. 1998, \pasp, 110, 761

\bibitem[{{Field} \& {Steigman}(1971)}]{Field71}
{Field}, G.~B., \& {Steigman}, G. 1971, \apj, 166, 59

\bibitem[{{Fox} {et~al.}(2007){Fox}, {Petitjean}, {Ledoux}, \&
  {Srianand}}]{Fox07}
{Fox}, A.~J., {Petitjean}, P., {Ledoux}, C., \& {Srianand}, R. 2007, \apjl,
  668, L15

\bibitem[{{Fumagalli} {et~al.}(2011){Fumagalli}, {Prochaska}, {Kasen}, {Dekel},
  {Ceverino}, \& {Primack}}]{Fumagalli11}
{Fumagalli}, M., {Prochaska}, J.~X., {Kasen}, D., {Dekel}, A., {Ceverino}, D.,
  \& {Primack}, J.~R. 2011, \mnras, 418, 1796

\bibitem[{{Goerdt} {et~al.}(2012){Goerdt}, {Dekel}, {Sternberg}, {Gnat}, \&
  {Ceverino}}]{Goerdt12}
{Goerdt}, T., {Dekel}, A., {Sternberg}, A., {Gnat}, O., \& {Ceverino}, D. 2012,
  \mnras, 424, 2292

\bibitem[{{Haardt} \& {Madau}(2012)}]{Haardt12}
{Haardt}, F., \& {Madau}, P. 2012, \apj, 746, 125

\bibitem[{{Jenkins}(2009)}]{Jenkins09}
{Jenkins}, E.~B. 2009, \apj, 700, 1299

\bibitem[{{Kacprzak} {et~al.}(2012){Kacprzak}, {Churchill}, {Steidel},
  {Spitler}, \& {Holtzman}}]{Kacprzak12}
{Kacprzak}, G.~G., {Churchill}, C.~W., {Steidel}, C.~C., {Spitler}, L.~R., \&
  {Holtzman}, J.~A. 2012, \mnras, 427, 3029

\bibitem[{{Keisler} {et~al.}(2011){Keisler}, {Reichardt}, {Aird}, {Benson},
  {Bleem}, {Carlstrom}, {Chang}, {Cho}, {Crawford}, {Crites}, {de Haan},
  {Dobbs}, {Dudley}, {George}, {Halverson}, {Holder}, {Holzapfel}, {Hoover},
  {Hou}, {Hrubes}, {Joy}, {Knox}, {Lee}, {Leitch}, {Lueker}, {Luong-Van},
  {McMahon}, {Mehl}, {Meyer}, {Millea}, {Mohr}, {Montroy}, {Natoli}, {Padin},
  {Plagge}, {Pryke}, {Ruhl}, {Schaffer}, {Shaw}, {Shirokoff}, {Spieler},
  {Staniszewski}, {Stark}, {Story}, {van Engelen}, {Vanderlinde}, {Vieira},
  {Williamson}, \& {Zahn}}]{Keisler11}
{Keisler}, R., {Reichardt}, C.~L., {Aird}, K.~A., {Benson}, B.~A., {Bleem},
  L.~E., {Carlstrom}, J.~E., {Chang}, C.~L., {Cho}, H.~M., {Crawford}, T.~M.,
  {Crites}, A.~T., {de Haan}, T., {Dobbs}, M.~A., {Dudley}, J., {George},
  E.~M., {Halverson}, N.~W., {Holder}, G.~P., {Holzapfel}, W.~L., {Hoover}, S.,
  {Hou}, Z., {Hrubes}, J.~D., {Joy}, M., {Knox}, L., {Lee}, A.~T., {Leitch},
  E.~M., {Lueker}, M., {Luong-Van}, D., {McMahon}, J.~J., {Mehl}, J., {Meyer},
  S.~S., {Millea}, M., {Mohr}, J.~J., {Montroy}, T.~E., {Natoli}, T., {Padin},
  S., {Plagge}, T., {Pryke}, C., {Ruhl}, J.~E., {Schaffer}, K.~K., {Shaw}, L.,
  {Shirokoff}, E., {Spieler}, H.~G., {Staniszewski}, Z., {Stark}, A.~A.,
  {Story}, K., {van Engelen}, A., {Vanderlinde}, K., {Vieira}, J.~D.,
  {Williamson}, R., \& {Zahn}, O. 2011, \apj, 743, 28

\bibitem[{{Kere{\v s}} {et~al.}(2005){Kere{\v s}}, {Katz}, {Weinberg}, \&
  {Dav{\'e}}}]{Keres05}
{Kere{\v s}}, D., {Katz}, N., {Weinberg}, D.~H., \& {Dav{\'e}}, R. 2005,
  \mnras, 363, 2

\bibitem[{{Kimm} {et~al.}(2011){Kimm}, {Slyz}, {Devriendt}, \&
  {Pichon}}]{Kimm11}
{Kimm}, T., {Slyz}, A., {Devriendt}, J., \& {Pichon}, C. 2011, \mnras, 413, L51

\bibitem[{{Krogager} {et~al.}(2012){Krogager}, {Fynbo}, {M{\o}ller},
    {Ledoux}, {Noterdaeme}, {Christensen}, {Milvang-Jensen}, \&
    {Sparre}}]{Krogager12} {Krogager}, J.-K., {Fynbo}, J.~P.~U.,
  {M{\o}ller}, P., {Ledoux}, C., {Noterdaeme}, P., {Christensen}, L.,
  {Milvang-Jensen}, B. \& {Sparre}, M., 2012, \mnras, 424, L1

\bibitem[{{Lehner} {et~al.}(2013){Lehner}, {Howk}, {Tripp}, {Tumlinson},
  {Prochaska}, {O'Meara}, {Thom}, {Werk}, {Fox}, \& {Ribaudo}}]{Lehner13}
{Lehner}, N., {Howk}, J.~C., {Tripp}, T.~M., {Tumlinson}, J., {Prochaska},
  J.~X., {O'Meara}, J.~M., {Thom}, C., {Werk}, J.~K., {Fox}, A.~J., \&
  {Ribaudo}, J. 2013, \apj, 770, 138

\bibitem[{{Nelson} {et~al.}(2013){Nelson}, {Vogelsberger}, {Genel}, {Sijacki},
  {Kere{\v s}}, {Springel}, \& {Hernquist}}]{Nelson13}
{Nelson}, D., {Vogelsberger}, M., {Genel}, S., {Sijacki}, D., {Kere{\v s}}, D.,
  {Springel}, V., \& {Hernquist}, L. 2013, \mnras, 429, 3353

\bibitem[{{Oppenheimer} {et~al.}(2010){Oppenheimer}, {Dav{\'e}}, {Kere{\v s}},
  {Fardal}, {Katz}, {Kollmeier}, \& {Weinberg}}]{Oppenheimer10}
{Oppenheimer}, B.~D., {Dav{\'e}}, R., {Kere{\v s}}, D., {Fardal}, M., {Katz},
  N., {Kollmeier}, J.~A., \& {Weinberg}, D.~H. 2010, \mnras, 406, 2325

\bibitem[{{Pettini} \& {Cooke}(2012)}]{Pettini12}
{Pettini}, M., \& {Cooke}, R. 2012, \mnras, 425, 2477

\bibitem[{{Pettini} {et~al.}(2002){Pettini}, {Rix}, {Steidel}, {Adelberger},
  {Hunt}, \& {Shapley}}]{Pettini02_cB58}
{Pettini}, M., {Rix}, S.~A., {Steidel}, C.~C., {Adelberger}, K.~L., {Hunt},
  M.~P., \& {Shapley}, A.~E. 2002, \apj, 569, 742

\bibitem[{{Pettini} {et~al.}(2001){Pettini}, {Shapley}, {Steidel}, {Cuby},
  {Dickinson}, {Moorwood}, {Adelberger}, \& {Giavalisco}}]{Pettini01_LBG}
{Pettini}, M., {Shapley}, A.~E., {Steidel}, C.~C., {Cuby}, J.-G., {Dickinson},
  M., {Moorwood}, A.~F.~M., {Adelberger}, K.~L., \& {Giavalisco}, M. 2001,
  \apj, 554, 981

\bibitem[{{Prochaska} {et~al.}(2002){Prochaska}, {Henry}, {O'Meara}, {Tytler},
  {Wolfe}, {Kirkman}, {Lubin}, \& {Suzuki}}]{Prochaska02a}
{Prochaska}, J.~X., {Henry}, R.~B.~C., {O'Meara}, J.~M., {Tytler}, D., {Wolfe},
  A.~M., {Kirkman}, D., {Lubin}, D., \& {Suzuki}, N. 2002, \pasp, 114, 933

\bibitem[{{Prochter} {et~al.}(2010){Prochter}, {Prochaska}, {O'Meara},
  {Burles}, \& {Bernstein}}]{Prochter10}
{Prochter}, G.~E., {Prochaska}, J.~X., {O'Meara}, J.~M., {Burles}, S., \&
  {Bernstein}, R.~A. 2010, \apj, 708, 1221

\bibitem[{{Rakic} {et~al.}(2011){Rakic}, {Schaye}, {Steidel}, \&
  {Rudie}}]{Rakic11}
{Rakic}, O., {Schaye}, J., {Steidel}, C.~C., \& {Rudie}, G.~C. 2011, \mnras,
  414, 3265

\bibitem[{{Ribaudo} {et~al.}(2011){Ribaudo}, {Lehner}, {Howk}, {Werk}, {Tripp},
  {Prochaska}, {Meiring}, \& {Tumlinson}}]{Ribaudo11}
{Ribaudo}, J., {Lehner}, N., {Howk}, J.~C., {Werk}, J.~K., {Tripp}, T.~M.,
  {Prochaska}, J.~X., {Meiring}, J.~D., \& {Tumlinson}, J. 2011, \apj, 743, 207

\bibitem[{{Rolleston} {et~al.}(2003){Rolleston}, {Venn}, {Tolstoy}, \&
  {Dufton}}]{Rolleston03}
{Rolleston}, W.~R.~J., {Venn}, K., {Tolstoy}, E., \& {Dufton}, P.~L. 2003,
  \aap, 400, 21

\bibitem[{{Rudie} {et~al.}(2012){Rudie}, {Steidel}, {Trainor}, {Rakic},
  {Bogosavljevi{\'c}}, {Pettini}, {Reddy}, {Shapley}, {Erb}, \&
  {Law}}]{Rudie12}
{Rudie}, G.~C., {Steidel}, C.~C., {Trainor}, R.~F., {Rakic}, O.,
  {Bogosavljevi{\'c}}, M., {Pettini}, M., {Reddy}, N., {Shapley}, A.~E., {Erb},
  D.~K., \& {Law}, D.~R. 2012, \apj, 750, 67

\bibitem[{{Rudie}(2013){Rudie}}]{Rudie13}
{Rudie}, G.~C. 2013, PhD thesis, CaltechTHESIS:05202013-155707736

\bibitem[{{Savage} \& {Sembach}(1991)}]{Savage91}
{Savage}, B.~D., \& {Sembach}, K.~R. 1991, \apj, 379, 245

\bibitem[{{Schaye} {et~al.}(2003){Schaye}, {Aguirre}, {Kim}, {Theuns}, {Rauch},
  \& {Sargent}}]{Schaye03}
{Schaye}, J., {Aguirre}, A., {Kim}, T.-S., {Theuns}, T., {Rauch}, M., \&
  {Sargent}, W.~L.~W. 2003, \apj, 596, 768

\bibitem[{{Shen} {et~al.}(2013){Shen}, {Madau}, {Guedes}, {Mayer}, {Prochaska},
  \& {Wadsley}}]{Shen13}
{Shen}, S., {Madau}, P., {Guedes}, J., {Mayer}, L., {Prochaska}, J.~X., \&
  {Wadsley}, J. 2013, \apj, 765, 89

\bibitem[{{Simcoe} {et~al.}(2002){Simcoe}, {Sargent}, \& {Rauch}}]{Simcoe02}
{Simcoe}, R.~A., {Sargent}, W.~L.~W., \& {Rauch}, M. 2002, \apj, 578, 737

\bibitem[{{Simcoe} {et~al.}(2004){Simcoe}, {Sargent}, \& {Rauch}}]{Simcoe04}
{Simcoe}, R.~A., {Sargent}, W.~L.~W., \& {Rauch}, M. 2004, \apj, 606, 92

\bibitem[{{Simcoe} {et~al.}(2006){Simcoe}, {Sargent}, {Rauch}, \&
  {Becker}}]{Simcoe06}
{Simcoe}, R.~A., {Sargent}, W.~L.~W., {Rauch}, M., \& {Becker}, G. 2006, \apj,
  637, 648

\bibitem[{{Steidel}(1990)}]{Steidel90_LLS}
{Steidel}, C.~C. 1990, \apjs, 74, 37

\bibitem[{{Steidel} {et~al.}(2010){Steidel}, {Erb}, {Shapley}, {Pettini},
  {Reddy}, {Bogosavljevi{\'c}}, {Rudie}, \& {Rakic}}]{Steidel10}
{Steidel}, C.~C., {Erb}, D.~K., {Shapley}, A.~E., {Pettini}, M., {Reddy}, N.,
  {Bogosavljevi{\'c}}, M., {Rudie}, G.~C., \& {Rakic}, O. 2010, \apj, 717, 289

\bibitem[{{Stewart} {et~al.}(2011){Stewart}, {Kaufmann}, {Bullock}, {Barton},
  {Maller}, {Diemand}, \& {Wadsley}}]{Stewart11}
{Stewart}, K.~R., {Kaufmann}, T., {Bullock}, J.~S., {Barton}, E.~J., {Maller},
  A.~H., {Diemand}, J., \& {Wadsley}, J. 2011, \apj, 738, 39

\bibitem[{{Stinson} {et~al.}(2013){Stinson}, {Brook}, {Macci{\`o}}, {Wadsley},
  {Quinn}, \& {Couchman}}]{Stinson13_EarlyFeedback}
{Stinson}, G.~S., {Brook}, C., {Macci{\`o}}, A.~V., {Wadsley}, J., {Quinn},
  T.~R., \& {Couchman}, H.~M.~P. 2013, \mnras, 428, 129

\bibitem[{{Tripp} {et~al.}(2005){Tripp}, {Jenkins}, {Bowen}, {Prochaska},
  {Aracil}, \& {Ganguly}}]{Tripp05}
{Tripp}, T.~M., {Jenkins}, E.~B., {Bowen}, D.~V., {Prochaska}, J.~X., {Aracil},
  B., \& {Ganguly}, R. 2005, \apj, 619, 714

\bibitem[{{van de Voort} {et~al.}(2012){van de Voort}, {Schaye}, {Altay}, \&
  {Theuns}}]{vandeVoort12_dla}
{van de Voort}, F., {Schaye}, J., {Altay}, G., \& {Theuns}, T. 2012, \mnras,
  421, 2809

\bibitem[{{Webb} {et~al.}(1991){Webb}, {Carswell}, {Irwin}, \&
  {Penston}}]{Webb91}
{Webb}, J.~K., {Carswell}, R.~F., {Irwin}, M.~J., \& {Penston}, M.~V. 1991,
  \mnras, 250, 657

\end{thebibliography}
\end{document}